# Li$_{4-x}$Ge$_{1-x}$P$_x$O$_4$ a potential solid-state electrolyte for all-oxide microbatteries

E. Gilardi,[1,2,*] G. Materzanini,[2,3,*] L. Kahle,[2,3] M. Döbeli,[4] S. Lacey,[2,5] X. Cheng,[1,2] N. Marzari,[2,3] D. Pergolesi,[1,2,5] A. Hintennach,[6] T. Lippert,[1,7]

[1] Research with Neutrons and Muons (NUM), Paul Scherrer Institute, 5232 Villigen PSI, CH

[2] National Centre for Computational Design and Discovery of Novel Materials (MARVEL), 1015 Lausanne, Switzerland

[3] Theory and Simulations of Materials (THEOS) École Polytechnique Fédérale de Lausanne, 1015 Lausanne, Switzerland

[4] Ion Beam Physics ETH Zürich CH-8093 Zürich, Switzerland

[5] Energy and Environment Research Division (ENE), Paul Scherrer Institute, 5232 Villigen PSI, CH

[6] Daimler AG, 70546 Stuttgart, Germany

[7] Department of Chemistry and Applied Biosciences, Laboratory of Inorganic Chemistry Vladimir-Prelog-Weg 1-5/10, ETH Zürich, 8093 Zürich, Switzerland

## Abstract

Solid-state electrolytes for Li-ion batteries are attracting growing interest as they allow building safer batteries, also using lithium metal anodes. Here we studied a compound in the lithium superionic conductor (LISICON) family, i.e. Li$_{4-x}$Ge$_{1-x}$P$_x$O$_4$ (LGPO). Thin films were deposited *via* pulsed laser deposition and their electrical properties were compared with ceramic pellets. A detailed characterization of the microstructure shows that thin films can be deposited fully crystalline at higher temperatures but also partially amorphous at room temperature. The conductivity is not strongly influenced by the presence of grain boundaries, exposure to air or lithium deficiencies. First-principles molecular dynamics simulations were employed to calculate the lithium ion diffusion profile and the conductivity at various temperatures of the ideal LGPO crystal. Simulations gives the upper limit of conductivity for a defect free crystal, which is in the range of 10$^{-2}$ S cm$^{-1}$ at 300 °C The ease of thin film fabrication, the room-temperature Li-ion conductivity in the range of a few μS cm$^{-1}$ make LGPO very appealing electrolyte material for thin film all–solid-state all-oxide microbatteries.

## Introduction

Lithium ion batteries are nowadays among the most widely used technology for energy storage. Conventional batteries use liquid organic solvents as the electrolyte, which are corrosive and flammable, therefore posing important safety issues. The risk of leaks precludes the potential use of these batteries for biomedical implants. A liquid electrolyte imposes also severe restrictions to the possible miniaturization of the battery, which is needed for on-chip power sources for portable/wearable applications. Also, Li metal dendrites can easily grow into a liquid electrolyte, short circuiting the electrodes.[1-2]

The use of solid-state electrolytes (SSE) would solve or mitigate these issues, as SSE are not flammable, cannot leak, and can in principle be grown as nanometric thin films for battery miniaturization. Many SSE are stable within a large electrochemical voltage window.[3-4] Finally, SSE hinder the formation of dendrites, although when metallic lithium anode is employed, the problem cannot be completely avoided.[5]

Inorganic oxides SSE are less conductive at room temperature than sulfides or halides, but generally more stable in contact with electrodes as well as when exposed to air. [3]

Different oxides show promising properties for possible use as SSE in batteries, for instance $Li_7La_3Zr_2O_{12}$ (LLZO) and $Li_{3x}La_{2/3-x}TiO_3$ (LLTO).[6-7] These materials offer lithium ion conductivities in the order of a few mS cm$^{-1}$ at room temperature. Also, ionic conductors in the LISICON family, with a composition of $Li_{4\pm x}M_{1-x}Y_xO_4$ (M = Ge, Si; Y = P, Al, Zn, Ge, Ga, Sb) are well known solid-state electrolytes, with a conductivity at room temperature in the range between a few and a few tens of μS cm$^{-1}$.[8-10] Recent works highlighted that the conductivity can be further improved exploiting the "mixed polyanion effect": the introduction of different cations causes a disordered redistribution of the anionic tetrahedra, which facilitates the lithium ion migration.[11-12]

Compounds in the LISICON family are also particularly interesting for the growth of thin films, as they can be deposited easily by sputtering [13-15] and pulsed laser deposition (PLD).[16] Thin films of LISICON materials reported in the literature are generally amorphous, which further enhances the lithium ion conductivity.[13]

On the contrary, the growth of thin films of other, much more conductive, oxides, such as LLZO and LLTO, turns out to be extremely difficult.[17-20] Different reports can be found in the literature on silicates,[15-16, 21] but nothing has been published to the best of our knowledge on germanate materials.

Here we report the deposition via PLD of thin films of lithium germanium phosphate, with the general formula $Li_{4-x}Ge_{1-x}P_xO_4$ (LGPO), a solid solution of γ-$Li_3PO_4$ and $Li_4GeO_4$.[22-23] Different deposition temperatures led to films with different crystalline quality and chemical composition. These features were then correlated to the electrical properties.

First-principles molecular dynamics simulations were then employed to calculate the Li-ion diffusion profile and diffusivity at different temperatures in an ideal (defect-free) LGPO crystal.

**Methods**

**Powder synthesis and characterization**

$Li_{3.2}P_{0.80}Ge_{0.20}O_4$ (LGPO) powders were

synthesized *via* the solid-state synthesis method using lithium carbonate ($Li_2CO_3$), germanium oxide ($GeO_2$) and lithium phosphate ($Li_3PO_4$). Lithium carbonate and germanium oxide were mixed in stoichiometric amount and grounded using an agate mortar, and the resulting powders were pressed uniaxially with 0.5 GPa. The pressed powders were sintered at 800 °C for 8 hours. Resulting $Li_4GeO_4$ and $Li_3PO_4$ powders were then mixed and pressed uniaxially with 0.5 GPa. The powders were then placed in a tubular furnace with constant oxygen flux at 900 °C for 12 hours. The resulting pellets were 13 mm in diameter with a thickness of 2 mm. The same procedure was repeated for sintering ceramic LGPO pellets used for conductivity measurements and as target for thin film deposition. The crystal structure of the synthesized LGPO powders were characterized by X-ray diffraction (Siemens D500 diffractometer with Cu-$K_\alpha$ radiation).

**Thin film deposition**

Thin films of LGPO were deposited *via* pulsed laser deposition (PLD) with a KrF excimer laser, wavelength 248 nm, laser fluence of 2.1 J/cm$^2$, repetition rate of 10 Hz. To compensate the expected Li loss (a problem often encountered during the growth of Li-containing thin films) a ceramic target with 10 mol% excess Li (as $Li_2O$) was used. The distance between target and substrate during deposition was 60 mm. Thin films were deposited on single crystal (100)-oriented MgO substrate. Silver paste was used to provide the thermal contact between the substrate and the heating stage in the vacuum chamber. The oxygen pressure was set to 0.01 mbar during the deposition. Films were deposited at 500 °C and at room temperature at the same oxygen background pressure.

**Thin film characterization**

The crystal structure of the films was characterized by X-ray diffraction (out of plane ω/2θ scan) with a Seifert diffractometer with monochromatic Cu-$K_{\alpha 1}$ radiation. The thin films deposited at low temperature were characterized also by 2θ scan. While ω was fixed at 1°, 2θ was scanned between 10° and 65°.

For the in-plane (along the direction of the substrate surface) electrical characterization two strip-shaped gold electrodes, 100 nm thick, were sputtered on the surface of the films.

Electrical characterization was performed by impedance spectroscopy (IS) in Ar flow, between 2 MHz and 1 Hz, applying an excitation of 1 V. Ag paste and Au wires were used to connect the electrodes to the measurement cell. A Solatron 1260 gain-phase analyzer and the software Zplot were used for the impedance measurement. The software Zview was used to fit the complex impedance plane plots. Thin films deposited at high temperature were equilibrated for 2 hours at 430 °C. The temperature was then decreased in steps of 20 °C and the samples were equilibrated for 30 minutes at each temperature step before acquiring a complex impedance plot. Thin films deposited at room temperature were equilibrated 24 hours at 25 °C and data was acquired while increasing the temperature up to 480 °C. The conductivity was also measured decreasing the temperature following the same procedure as described above.

Rutherford Back Scattering was performed with a 2 MeV 4He beam and a silicon PIN

diode detector under 168°. Data were analyzed by the RUMP code. [24] Elastic Recoil Detection Analysis with a 13 MeV 127I primary ion beam was primarily used to determine the lithium-to-oxygen ratio. Recoiling ions were detected by the combination of a time-of-flight spectrometer with a gas ionization chamber. The acquired spectral data were analyzed by a custom software.

**First-Principles molecular dynamics**

To examine Li-ion diffusivity in LGPO on the atomic scale in a perfect crystal with a composition of $Li_{3.33}Ge_{0.33}P_{0.67}O_4$, extended first-principles molecular dynamics (FPMD) simulations were performed. The Car-Parrinello (CPMD) scheme was used, based on Kohn-Sham density-functional theory (DFT) [25-26] in the plane-wave pseudopotential formalism, [27] as implemented in the cp code of the Quantum ESPRESSO distribution (http://www.quantum-espresso.org/).[28] A plane-waves cutoff of 50 Ry for the wavefunctions and 400 Ry for the electron density and ultrasoft pseudopotentials from Standard Solid State Pseudopotential (SSSP) Efficiency library 1.0 [29] (GBRV [30] for Li and O, PSLib [31] for Ge and P) were employed, with Brillouin-zone integration at the Gamma point. The exchange-correlation functional was PBE. [32]

CPMD simulations, with a time step of 4 a.u. and an electronic mass of 500 a.u., were performed for 200 – 500 ps in the fixed simulation (1x1x3) supercell defined by the experimental crystallographic data [23] at six temperatures ranging between 600 K and 1400 K (600, 720, 900, 1000, 1100, 1400 K). To sample the NVT ensemble, temperature was controlled via a Nosé-Hoover thermostat for each atomic species.[33-34]

From the NVT trajectories, the Li-ion positions $\{R_i(t)\}$ were extracted and the Li-ion probability density $n_{Li}(\vec{r})$ was calculated as:

(1)
$$n_{Li}(\vec{r}) = (\sigma\sqrt{2\pi})^{-3} \left\langle \sum_{l}^{N_{Li}} e^{-\frac{|\vec{r}-\vec{R}_l(t)|^2}{2\sigma^2}} \right\rangle_t .$$

The normal distribution was used to account for finite statistics, setting the deviation to σ = 0.3 Å, and the summation was performed on a grid of 0.1 Å$^{-1}$.

In addition, we analyzed the trajectory using the SITATOR package for an unsupervised analysis of the diffusive pathways in the system. [35]

The tracer diffusion matrix $D^{Li}$ was computed from the NVT trajectories via the mean square displacement (MSD) of lithium, according to the Einstein relation:

(2)
$$D_{ij}^{Li} = \lim_{\tau\to\infty} \frac{1}{2\tau} \frac{1}{N_{Li}} \sum_{l}^{N_{Li}} \left\langle \left[ R_i^l(t+\tau) - R_i^l(t) \right]\left[ R_j^l(t+\tau) - R_j^l(t) \right] \right\rangle_t =$$
$$= \lim_{\tau\to\infty} \frac{1}{2\tau} MSD_{ij}(\tau) ,$$

where $R_i^l(t)$ is the i-component ($i = x, y, z$) of the $l-th$ Li-ion position at time t, and the angular brackets $\langle\ \rangle_t$ define a time average over the starting times t, that, following the ergodic hypothesis, we assume corresponding to an ensemble average. We computed this average, and its statistical error, by means of a block analysis.[36] The tracer diffusion coefficient

$D_{tr}^{Li}$ was calculated from the $D^{Li}$ matrix as $1/3\ Tr(D^{Li})$, or directly from the formula

(3)
$$D_{tr}^{Li} = \lim_{\tau \to \infty} \frac{1}{6\tau} \frac{1}{N_{Li}} \sum_{l}^{N_{Li}} \left\langle \left| \vec{R}^l(t+\tau) - \vec{R}^l(t) \right|^2 \right\rangle_t =$$
$$= \lim_{\tau \to \infty} \frac{1}{6\tau} MSD(\tau).$$

From the tracer diffusion coefficient at each temperature, the ionic conductivity was extracted *via* the Nernst-Einstein equation:

(4)
$$\sigma(T) = \frac{N_{Li} e^2}{V k_B T} D_{tr}^{Li}(T)$$

where $N_{Li}/V$ is the Li-ions density and $e$ is the elementary charge. We assumed an average Born effective charge for Li of +1.[37]

$k_B$ is the Boltzmann constant and $T$ the absolute temperature.

**Results and Discussion**

**LGPO ceramic pellets**

LGPO ceramic pellets, with a nominal composition of $Li_{3.2}Ge_{0.2}P_{0.8}O_4$ were prepared by solid-state synthesis, then pressed and sintered. The density of the pellets was 2.20 g cm$^{-3}$, corresponding to 85 % of the theoretical density. The XRD analysis and comparison with the ICSD reference (Figure 1a) indicates the formation of the expected LGPO orthorhombic structure (space group P n m a). Three minor diffraction peaks corresponding to lithium carbonate and lithium phosphate were also detected.

The electrical conductivity was measured by impedance spectroscopy (IS) with gold (Li-ion blocking) electrodes. Two representative Nyquist plots, used to calculate the total resistance, are reported in Figure 1b. With our samples and selected experimental conditions, it was not possible to clearly distinguish the grain boundary and grain interior contributions to the conductivity. Therefore, only the total resistance could be reliably measured.

**LGPO thin films**

LGPO thin films were grown by pulsed laser deposition (PLD) on MgO (100)-oriented single crystal substrates. A ceramic pellet fabricated in our laboratory was used as target for the ablation process.

In the case of complex multi-element materials, containing elements with very different atomic masses, the target-to-film compositional transfer can be a severe issue.[38-39] Many factors, other than the target composition, can influence the film elemental content. Due to the interaction of the ablated species with the gaseous environment during the time of flight towards the substrate, pulsed laser deposited films are typically enriched with the heavier elements at the expense of the lighters.[40] This is why it is very difficult to grow films with the desired Li content. To partially compensate for the Li loss, 10 % mol excess Li was added to the target as $Li_2O$. Films with thickness between 150 nm and 300 nm were deposited in an oxygen background pressure of 0.01 mbar. The samples were prepared at room temperature and at 500 °C to probe the effects of temperature on composition, crystallinity, and transport properties.

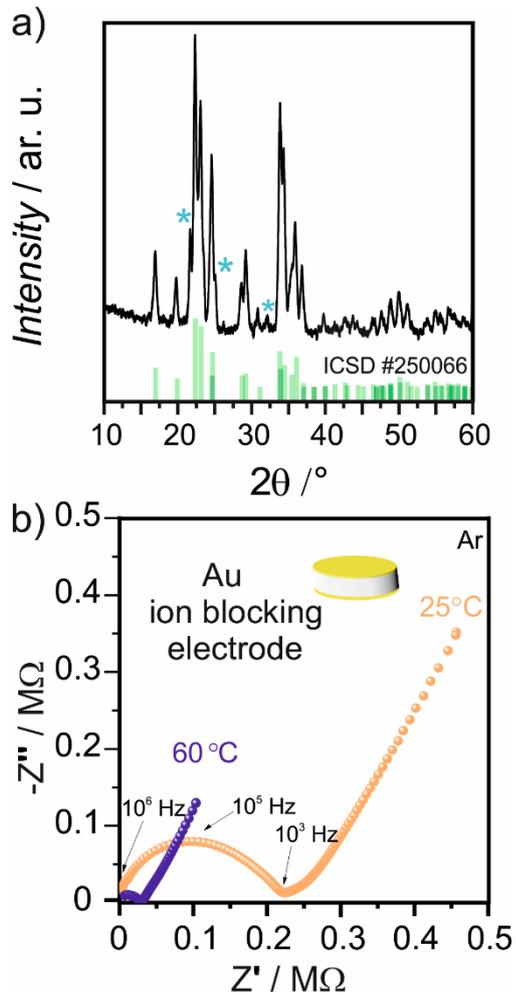

**Figure 1** XRD and IS characterization of LGPO (Li$_{3.2}$Ge$_{0.2}$P$_{0.8}$O$_4$) pellet. a) X-ray diffraction pattern of LGPO ceramic pellet after sintering at 900 °C, 12 h. In green the comparison with the ICSD reference 250066. The blue stars indicate peaks originating from secondary phases. b) Nyquist plots obtained at temperatures of 60 °C and 25 °C in Ar gas, from the same ceramic pellet with sputtered gold electrodes. In the inset: a scheme of the electrode configuration is shown.

In Figure 2a-I the XRD analysis of a film grown at high temperature is presented. Except minor peaks stemming from secondary phases, all diffraction peaks can be ascribed to the orthorhombic structure of LGPO. The films appear to be polycrystalline, as expected, due to the absence of crystallographic matching between films and substrates.

Instead, no diffraction peaks can be seen in the plot of the 2θ/ω scan of the sample grown at room temperature (Figure 2a-II). For this sample, more can be learned from the plot of the 2θ scan shown in Figure 2b. This XRD pattern can be explained assuming a nanocrystalline film, which could also be partially amorphous. However, the presence of diffraction peaks clearly ascribable to the LGPO structure indicates a certain degree of crystallinity also for the films deposited at room temperature. This is noteworthy, because in the literature LISICON thin films deposited *via* PLD were reported to be amorphous, on the basis of the absence of diffraction peaks in the 2θ/ω scan.[16] Our results show that this is not always the case. An accurate analysis of the crystallinity is important, as complete film amorphization is known to improve the conductivity in LISICON materials.[13]

Figure 2c shows representative complex impedance plane plots acquired at the same temperature with different samples. The electrical resistance was measured in-plane. MgO is a very good electrical insulator, therefore no contribution to the resistance is expected from the substrate.

The Nyquist plots were fitted to the response of an equivalent circuit composed of a resistance (R) and a constant phase element (Q) in parallel (inset in Figure 2c). In the case of in-plane resistance measurements of thin films on substrates, the bulk and grain boundary contributions to the total resistance cannot be distinguished, due to the stray capacitance induced by the substrate and experimental set up.[41] Therefore, only the total conductivity can be deduced from the IS plot.

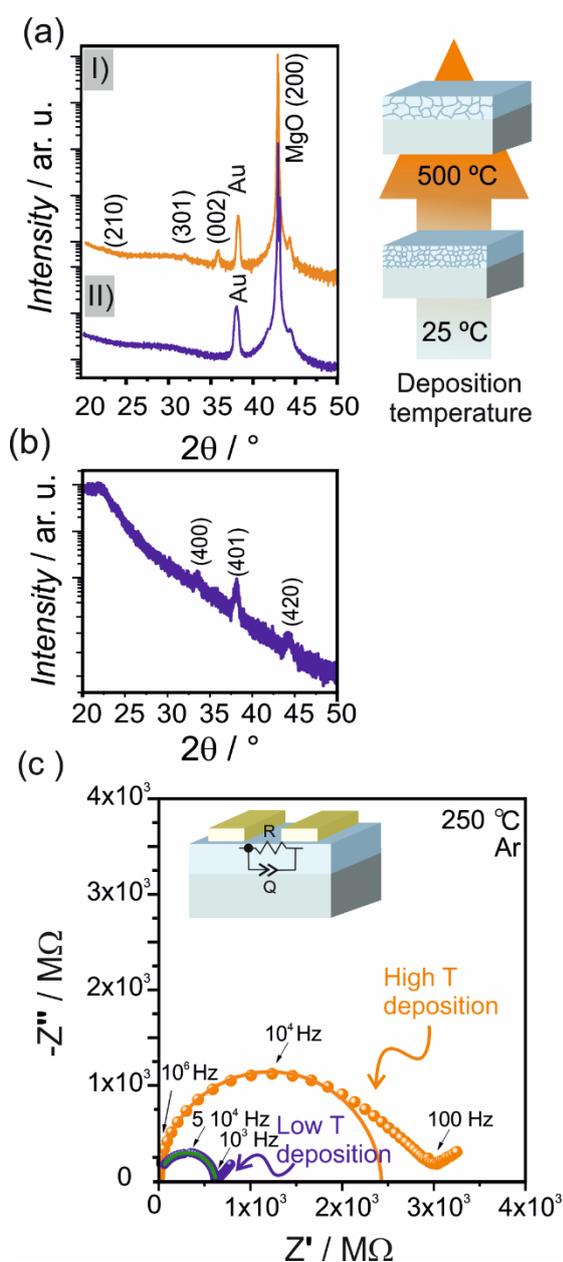

**Figure 2**
a) X-ray diffraction pattern (2θ/ω) of a thin film deposited at 500 °C (I) and thin film deposited at room temperature (II). The Miller indexes of the diffraction peaks of the LGPO film, gold electrodes and MgO substrate are reported. b) 2θ scan (ω kept constant at 1°) of the LGPO thin film deposited at room temperature. c) Nyquist plots measured at 250 °C on LGPO thin films deposited at different temperatures. The thin film deposited at 500 °C is 150 nm and the thin film deposited at 25 °C is 300 nm thick. Solid lines represent the plot obtained by fitting the data to the equivalent circuit shown in the inset.

According to Rutherford Back Scattering and Elastic Recoil Detection Analysis the composition of the thin films deposited at room temperature and 500 °C is $Li_{1.58\pm0.4}Ge_{0.45}P_{0.55}O_{3.27\pm0.5}$ and $Li_{2.7\pm0.3}Ge_{0.49}P_{0.5}O_{3.7\pm0.3}$, respectively. The experimental uncertainty of these measurements is about 4 % for the Ge and P content.

At both temperatures, a pronounced Li deficiency was observed, as expected. The two complementary ion beam-based compositional analyses also revealed the formation of carbonates. Lithium carbonate is a very common contaminant in Li-containing material when exposed to air. The presence of carbonates is particularly high in the film grown at room temperature, while for the film deposited at high temperature only a thin layer at the surface was detected. This observation suggests a much rougher and porous micromorphology of the LGPO films grown at room temperature, with respect to those grown at 500 °C. Roughness and porosity widen the surface area of the film, therefore allowing the formation of carbonates not only at the surface, as in the case of dense films. The formation of carbonate was taken into account for the calculation of the thin film compositions.

**LGPO ionic conductivity**

The conductivity of the LGPO ceramic pellets fabricated for this study is about 1.2 μS cm$^{-1}$ at room temperature, with an activation energy of about 0.53 eV. Figure 3 shows a representative conductivity Arrhenius plot for our samples in comparison with available data from the literature.

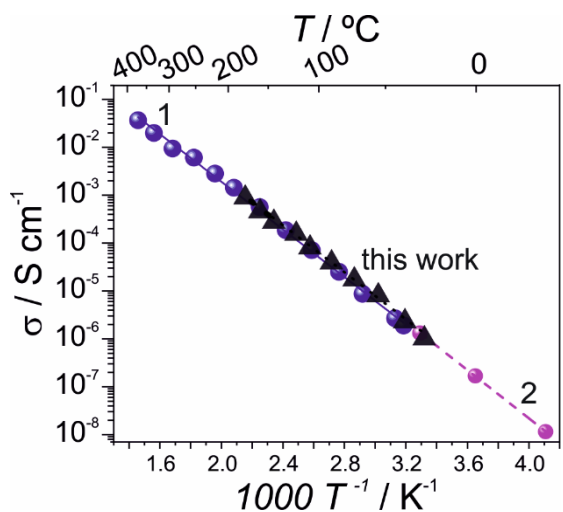

**Figure 3** Conductivity dependence on the inverse temperature of of LGPO ceramic pellets and single crystal. Our sample has the nominal composition $Li_{3.2}P_{0.8}Ge_{0.2}O_4$. Data of line 1 ($Li_{3.34}P_{0.66}Ge_{0.34}O_4$ single crystal) are taken from Ref.[22], data of line 2 ($Li_{3.2}P_{0.8}Ge_{0.2}O_4$ pellet) from Ref.[10]

Our measurements, obtained with pellets with a nominal composition of $Li_{3.2}Ge_{0.2}P_{0.8}O_4$ (x = 0.8), are in very good agreement with values reported in the literature for pellets with the same composition.[10, 42] A very similar conductivity was reported also for LGPO with x = 0.2, whereas the peak conductivity was found for x between 0.4 and 0.6 reaching values of about 10 μS cm$^{-1}$.[10] Another study reports a room-temperature conductivity of about 35 μS cm$^{-1}$ for LGPO pellets with x ≈ 0.25.[42] LGPO single crystal, with x = 0.66, showed a conductivity which is slightly lower but with the same activation energy. The value of 1.8 μS cm$^{-1}$ was reached at 40 °C.[22]

The very similar value of conductivity measured for single crystals[22] and sintered pellets[10, 43] suggests that the secondary phases in the sample, shown by XRD, do not affect significantly the transport properties.

In previous works, LGPO pellets with different Ge to P ratio (from 0.25 to 1.5) showed conductivities ranging from 1 to 10 μS cm$^{-1}$ at room temperature and the reported activation energies are all around 0.53 eV.[10, 43] The present work confirms these findings.

The correlation between chemical composition, morphology and conductivity seems to be not completely clear. It is however clear that the ionic conductivity near room temperature of LGPO with different compositions ranges between 1 and a few tens μS cm$^{-1}$.

Figure 4 compares the conductivity of one of the pellets fabricated for this work with the conductivity of LGPO thin films grown at room temperature and at 500 °C.

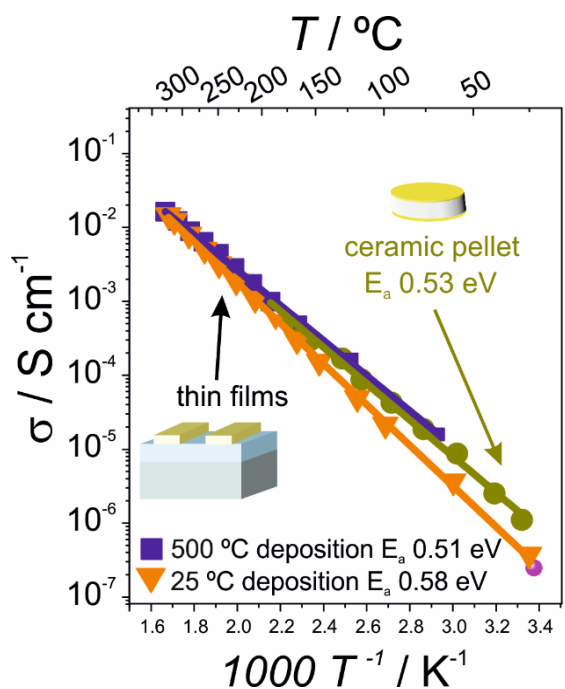

**Figure 4** Conductivity dependence on the inverse temperature of LGPO pellet ($Li_{3.2}Ge_{0.2}P_{0.8}O_4$) and thin films deposited at 25 and 500 °C. Composition of thin films from RBS and ERDA analysis: $Li_{1.58\pm0.4}Ge_{0.45}P_{0.55}O_{3.27\pm0.5}$ and $Li_{2.7\pm0.3}Ge_{0.49}P_{0.5}O_{3.7\pm0.3}$, respectively. The point (pink sphere) at 25 °C was measured for the thin film deposited at room temperature "as grown". After the complete thermal cycle, the conductivity did not show any significant change. The inset shows the electrode configuration used for the different samples.

LGPO thin films show electrical properties that are very similar to those of the pellets, independently of the deposition temperature. The film grown at room temperature shows an activation energy that is about 0.07 eV higher.

This range of conductivity overlaps with that reported in the literature for LiPON amorphous thin films (highest value 2.3 µS cm$^{-1}$ at room temperature).[44-45] LiPON is, so far, the only solid-state Li0-ion conductor that can be used as thin film electrolyte for microbatteries. The perovskite-type oxide LLTO and the garnet-type oxide LLZO, with room temperature conductivity in the range of mS cm$^{-1}$, require sintering temperatures well above 1000 °C for ceramic pellets and deposition temperatures well above 800 °C for thin film growth. This promotes Li evaporation and the stabilization of the $La_2Ti_2O_7$ and $La_2Zr_2O_7$ non-conductive phases that are thermodynamically favored. The conductivity of LLTO thin films reported so far in the literature is in the same range of LiPON and LGPO (but require a much higher processing temperature), while much lower conductivity has been reported for LLZO films.

An advantage of LISICON thin films is that the same conductivity of the stoichiometric compound can be reached already for low lithium content. For example, in case of solid solutions of lithium silicate – lithium phosphate ($Li_4SiO_4$ – $Li_3PO_4$), the conductivity strongly depends on the lithium stoichiometry only up to a certain threshold ( Li/(P+Si) ~ 1). Above this value, the conductivity becomes independent of the lithium content as the mobile lithium sites are saturated.[13] Assuming that LGPO behaves similarly to silicates, we can observe that in the film deposited at room temperature the measured lithium stoichiometry is lower compared to sample deposited at 500 °C. However, in both cases the ratio between Li and P+Ge contents is well above 1. Therefore, it is likely that other factors, such as different morphological features and/or the presence of $Li_2CO_3$, are affecting the ionic transport.

Considering the results of the XRD analysis in Figure 2a, these measurements confirm also for thin films what was deduced from the comparison of conductivities of sintered pellets and single crystals (Figure 3): the degree of crystallinity (average grain size, extent of grain boundaries) has a relatively small influence on the conducting properties.

**First-principles molecular dynamics simulations for bulk defect-free LGPO**

A 100 - atom (1 × 1 × 3) supercell was built starting from the single crystal unit cell for which crystallographic data are available in the literature (ICSD #250066 [23]). Stoichiometry of x = 0.66 was needed, in order to ensure a P to Ge ratio equal to 1:2.[22-23] This supercell, with a formula of $Li_{40}Ge_4P_8O_{48}$, is reported in Figure 5.

To understand Li-ion density and diffusion properties in an ideal (defect-free) LGPO crystal, we performed extended (200 ps – 500 ps) FPMD simulations on the above described periodic supercell in the NVT ensemble, between 300 °C and 1100 °C.

The isosurfaces of the Li-ion probability density at different temperatures, displayed in Figure 6 a and 6 b, indicate that the ideal LGPO crystal offers a three-dimensional isotropic conduction pathway for Li ions. This is consistent with the small degree of anisotropy shown by the Li-ion MSD along $x$, $y$, $z$, at the temperature of 1400 K (Figure 6 c). It is noteworthy that LGPO does not present unidimensional channels for the lithium ion conduction. Channel-blocking effects to the ionic transport can be ruled out and the diffusion is isotropic.

The absence of preferential conduction pathways for the ionic transport is an important advantage, since channel-blocking effects can lower significantly the diffusion and have detrimental consequences on the charge/discharge properties, when the material is used as electrolyte in batteries.[35, 44]

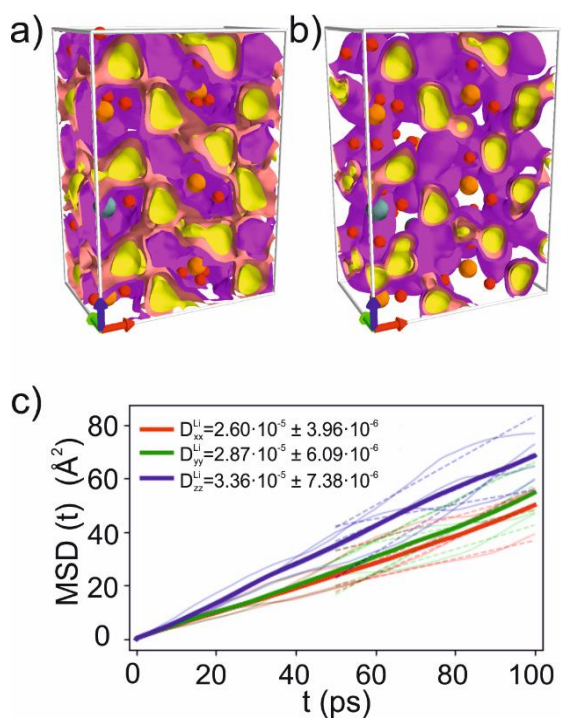

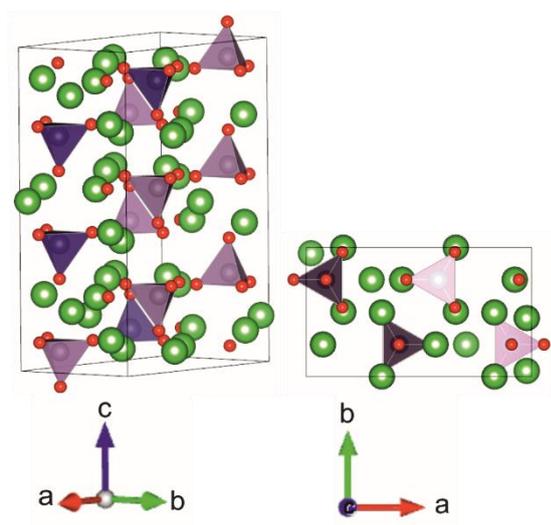

**Figure 5** Side and top view of the 100 - atom unit cell used for the simulations of defect-free LGPO crystal (from Ref. [23]). Li atoms are displayed in green, oxygen atoms in red, Ge and P atoms are at the center of the light and dark purple tetrahedra, respectively.

**Figure 6** The Li - ion probability density at a) 1400 K (1126 °C) and b) 600 K (327 °C), for three isovalues, 0.001 Å$^{-3}$, 0.01 Å$^{-3}$ and 0.1 Å$^{-3}$, as purple, salmon and yellow isosurfaces, respectively. The equilibrium positions of oxygen, phosphorus and germanium are shown as red, orange, and green spheres, respectively. c) x-, y- and z-resolved mean-squared displacement (MSD) for Li$^+$ as a function of time at 1400 K (1126 °C), as red, green, and blue solid lines, respectively, showing again isotropic diffusion. In the legends, the resulting diagonal elements of the diffusion matrix (and the error originating from the individual blocks, dashed lines) are reported.

Li-ion MSDs were used to calculate the ionic diffusion coefficients and the conductivity dependence on the temperature.

Figure 7 compares the computational values of Li-ion conductivity in LGPO with the experimental values previously described (high-temperature deposited thin film, x=0.5) obtained in this study.

The non-quantitative agreement between FPMD-calculated (0.37±0.04 eV) and IS-measured (0.51±0.01 eV) activation barriers is not new for ionic conductivities in solid-state electrolytes.[46-48] Possible reasons are the different time and length scales accessed through simulations (atomic level) and experiments (entire sample), and the absence of defects in the simulated ideal crystal, so that values calculated by FPMD may be considered as the upper limit for the conductivity in LGPO ($Li_{3.33}Ge_{0.33}P_{0.66}O_4$). Besides, going beyond the Li-ion tracer diffusion (i.e. calculating the charge diffusion coefficient [[34]]) would be desirable. However, our attempts to calculate the Haven ratio gave rise to charge diffusion coefficients with large statistical errors, that we don't report here, due to LGPO being a weak ionic conductor, so that longer trajectories would be needed for such calculations.[49] It is worthwhile to recall here that some of us have shown the Haven ratio not to play a significant role on the activation energy for the diffusion for the similar system LGPS.[34]

In turn, comparison with the experimental values, obtained in this work and reported in the literature, suggests that the transport properties are affected by the composition and presence of large-scale defects. Large-scale defects affect the conductivity measured by IS, while their effect is not included in the conductivity calculated by FPMD.

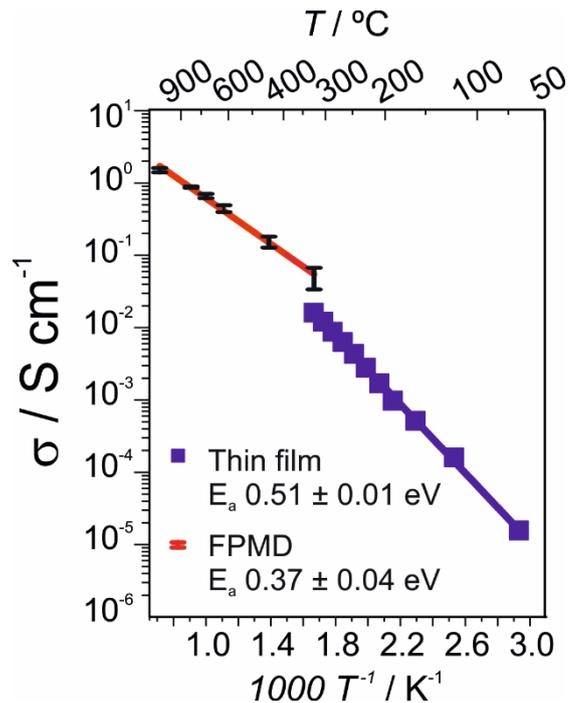

**Figure 7)** Li-ion conductivity Arrhenius plot for the simulated ideal LGPO (x=0.67) crystal (FPMD), compared with experimental data (high-temperature deposited thin film, x=0.5). The underestimation of the activation energy by simulations compared to the experimental values was previously ascribed to the different length scales analyzed by the techniques and difficulty to take 1D and 2D defects into account.[48]

**Conclusions**

In this work we synthesized and characterized pellets and thin films of $Li_{4-x}Ge_{1-x}P_xO_4$ (solid solution between γ-$Li_3PO_4$ and $Li_4GeO_4$) belonging to the ionic conductors in the LISICON family.

Pellets with nominal composition $Li_{3.2}Ge_{0.2}P_{0.8}O_4$ and a relative density of 85% show a conductivity and activation energy in agreement with previously published data.

Thin films were deposited *via* PLD at different temperatures. The compositional transfer from target to substrate during laser ablation is not stoichiometric, resulting in different ratios between Ge and P and the lithium content. Samples deposited at low temperature are deficient in lithium and show contaminations in the form of lithium carbonate. In addition, thin films deposited at low temperature are not completely amorphous as reported previously for similar materials, but appear to possess a nanocrystalline morphology.

Compared to the deposition at high temperature, the deposition at low temperature reduces the conductivity of the films by approximately a factor of 4 at room temperature. This is most probably due to the higher roughness and porosity, which favors the formation of carbonates throughout the film. Lithium carbonate is present as a contaminant only at the surface of the dense crystalline films without strongly affecting the conductivity after exposing the sample to air. .

From the characterization of pellet and thin films, it appears that the conductivity is almost independent of the microstructure and lithium content, if this is above the threshold of saturation of the mobile lithium sites.

Conductivities at different temperatures calculated by FPMD and experimental results are then compared. A clear, but expected difference exists between the short length scale conductivity calculated by FPMD and the conductivity measured on the whole sample by IS. This suggests that there is a blocking effect of large-scale defects that is not detected by the experimental characterization. Effects of the Li-ion collective diffusion (Haven ratio) were not taken into account here, due to the modest conductivity of this material, that would have required longer simulation times. At the same time, it was reported previously for similar systems that the Haven ration does not significantly affect the activation barrier. [34]

In conclusion, LGPO exhibits a number of very interesting features that are potentially important for a technological application as electrolyte in thin film batteries and fundamental research. In particular, the fabrication of thin films with the desired characteristics is particularly easy to achieve (also at room temperature).

It was previously reported that LGPO (with formula $Li_{3.33}Ge_{0.33}P_{0.66}O_4$) is stable between -0.5 and 7 V vs Li/Li$^+$, which make it suitable to be used also with high voltage electrodes.[43]

All these characteristics make LGPO a potentially interesting electrolyte material for all-solid-state all-oxide microbatteries.

AUTHOR INFORMATION

**Corresponding Author**

* elisa.gilardi@psi.ch

* giuliana.materzanini@epfl.ch

**Author Contributions**

The manuscript was written through contributions of all authors. All authors have given approval to the final version of the manuscript.

**Funding Sources**

This research was supported by the NCCR MARVEL, funded by the Swiss

**Notes**

EG, TL and DP would like to dedicate this work to their colleague, Andreas Hintennach, who unfortunately, died prematurely this year. He will be sorely missed and his personal, scientific and technical contributions will be irreplaceable.